\documentclass[article]{aa} 
\usepackage{graphicx}
\usepackage{txfonts}

\newcommand\simlt{\lower.5ex\hbox{$\; \buildrel < \over \sim \;$}}
\newcommand\simgt{\lower.5ex\hbox{$\; \buildrel > \over \sim \;$}}

\newcommand\rs[1]{_\mathrm{#1}}
\newcommand\zr{\rs{o}}

\newcommand\be{\begin{equation}}
\newcommand\ee{\end{equation}}
\newcommand\ba{\begin{eqnarray}}
\newcommand\ea{\end{eqnarray}}

\begin{document}

\title{Local Kelvin-Helmholtz instability and synchrotron modulation in Pulsar Wind Nebulae}
\author{
N. Bucciantini\inst{1} \and L. Del Zanna\inst{2} }
\institute{
Astronomy Department, University of California at Berkeley, 601 Campbell Hall, Berkeley, CA 94720-3411, USA \\
\email{niccolo@astron.berkeley.edu}
\and
Dipartimento di Astronomia e Scienza dello Spazio, Universit\`a di Firenze, L.go E.~Fermi 2, 50125 Firenze, Italy
}

\date{Received 7 November 2005 / Accepted 14 March 2006}

\abstract{We present here a series of numerical simulations of the development of Kelvin-Helmholtz instability in a relativistically hot plasma. The physical parameters in the unperturbed state are chosen to be representative of local conditions encountered in Pulsar Wind Nebulae (PWNe), with a main magnetic field perpendicular to a mildly relativistic shear layers. By using a numerical code for Relativistic MHD, we investigate the effect of an additional magnetic field component aligned with the shear velocity, and we follow the evolution of the instability to the saturation and turbulent regimes. Based on the resulting flow structure, we then compute synchrotron maps in order to evaluate the signature of Kelvin-Helmholtz instability on the emission and we investigate how the time scale and the amplitude of the synchrotron modulations depend on shear velocity and magnetic field. Finally we compare our results to the observed variable features in the Crab Nebula. We show that the Kelvin-Helmholtz instability cannot account for the wisps variability, but it might be responsible for the time dependent filamentary structure observed in the main torus.
}

\maketitle

\section{Introduction}
Pulsar Wind Nebulae (PWNe) arise from the confinement of the ultra-relativistic wind, powered by the pulsar spin-down energy, by the surrounding supernova ejecta, which are in free expansion in the early phases after the supernova event. The prototype of this class of objects is certainly the Crab Nebula. PWNe are characterized by continuous non thermal emission (synchrotron and inverse Compton), due to the injection of particles downstream of the termination shock (the discontinuity between the wind and the PWN) and to the presence of a magnetic field, which is basically the residual (toroidal) field carried by the wind, amplified after the termination shock (Kennel \& Coroniti \cite{kennel84}).

Recent 2D-axisymmetric Relativistic MHD (RMHD) simulations of PWNe (Komissarov \& Lyubarsky \cite{komissarov04}, Del Zanna et al. \cite{ldz04}), aimed at reproducing the main dynamical features of the inner structure of PWNe such as equatorial rings and polar jets, have shown that, due to the anisotropy in the wind energy flux (Bogovalov \cite{bogovalov01}), high velocity flow channels develop inside such objects. It has been suggested that the interface between such channels can be subject to Kelvin-Helmholtz (KH) instability (Begelman \cite{begelman99}), and that this might be at the origin of the time variability of some of the features present in the X-ray maps of the Crab Nebula, namely the wisps and possibly the filamentary structures in the main torus (Tanvir et al. \cite{tanvir97}, Hester et al. \cite{hester02}). However, at present, temporal and spatial resolution in global 2D simulations of PWNe is not sufficient to properly investigate the development of such instability and to follow its growth up to the non linear regime. 

To verify if KH instability can be at the origin of the observed time variability it is important to understand if the time scale for its growth and the signature on the emitted synchrotron radiation do agree with observations. Global 2D simulations also assume a purely toroidal magnetic field (no magnetic field component is aligned with the flow speed), while it is known that even a small component of poloidal field might substantially affect the growth and evolution of the perturbation (Miura \cite{miura84}, Malagoli et al. \cite{malagoli96}) leading to different observational signatures. For example, in a purely toroidal case the energy cascades toward large wavelengths, while if a poloidal field is present the cascade is toward small wavelengths. Synchrotron maps of PWNe seem in fact to suggest the possibility of a small scale turbulent magnetic field in order to explain features as the jet, whose synchrotron emission should be strongly suppressed in the case of a purely toroidal magnetic field (Komissarov \& Lyubarsky \cite{komissarov04}).

Let us briefly recall how relativity modifies the developement of the instability. In the case of a parallel magnetic field, relativistic flows tend to stability both in terms of the growth rate and range of the unstable velocity (the lower and upper cutoff in the Mach number close together. For mildly relativistic flows the lower cutoff shifts to lower values (destabilization at smaller speeds) while for ultrarelativistic flows instability is present only  for oblique propagation. In general a relativistically hot fluid has slightly higher growth rates, but deviations are important only very close to the stability threshold. In the case of a purely transverse magnetic field there is always a range of instability even for parallel propagation. Again the upper (and lower cutoff) at first decreases for mildly relativistic flows and then increases. As in the classic case secondary modes are found to develop for oblique propagation. The region for instability  of secondary modes overlaps with that of primary modes. It is also found for very oblique propagation in the ultrarelativistic case that the instability region splits and a narrow range of stable modes is found. 

Concerning the non linear phases, in general one expects that in a relativistically hot plasma both magnetic field and internal energy contribute to the inertia. This implies that even if $P >> \rho v^2$, the incompressible approximation fails. The consequence is that the pressure fluctuations due to the formation of large eddies, can be of the same order of the background pressure.  Concerning the fluctuations of the velocity components, it is found that the amplitude of the velocity perturbation in the direction parallel to the interface is smaller than that perpendicular to the interface. This is indeed a consequence of the relativistic dynamics.

Recently Bodo et al. (\cite{bodo04}) have shown that it is possible to reformulate the KH instability for relativistic unmagnetized fluid in an equivalent way to the non-relativistic case, by redefining the Mach number.  Given that the equations for transverse relativistic MHD have the same mathematical structure of the one for relativistic HD (Romero et al. \cite{romero05}), we can conclude that we do expect a similarity to hold also for transverse MHD.

Here we present the first numerical simulations of the growth of KH instability in a RMHD regime, in the simplified plane parallel geometry, which can be considered an extension of previous works by Miura (\cite{miura84}) and Malagoli et al. (\cite{malagoli96}), with which our simulations share many common points. KH instability has been widely studied by means of numerical simulations in the non relativistic regime, both magnetized and unmagnetized (e.g. Keppens et al. \cite{keppens99} and references therein). The recent development of numerical schemes for relativistic fluid dynamics and MHD allow us to finally extend such results to the relativistic regime. Recently numerical simulations were presented for unmagnetized high Lorentz factor jets (Perucho et al. \cite{perucho04}). Our simulations mainly focus on the subsonic or mildly supersonic regime typical of the flow inside PWNe, and take into account the possible presence of a poloidal magnetic field. Based on the numerical results for the flow structure we then compute synchrotron emission maps associated with the development of the instability, to be compared with observations of the Crab Nebula. We also evaluate the effect of the poloidal component on the polarization properties of the emitted radiation. 

\section{Numerical method and initial conditions}
Simulations were performed using the shock-capturing code for RMHD developed by Del Zanna et al. (\cite{ldz03}), to which the reader is referred for a detailed description of the algorithm. The scheme is particularly simple, since solvers based on characteristic waves are avoided in favor of central-type component-wise techniques (HLL solver). Density, momentum, energy, and a passive scalar that traces the mass function and allows us to evaluate the mixing, are evolved using a conservative approach, while the magnetic field is evolved via an ideal induction equation by following the Upwind Constrained Transport method (UCT, Londrillo \& Del Zanna \cite{londrillo04}), which guarantees that the solenoidal condition is satisfied up to machine accuracy if second order is employed for interpolation. We solve the RMHD equations in a 2D planar domain with Cartesian coordinates ($x,y$), where the shear layer is located at $y=0$.

Given the high Lorentz factor of the pulsar wind ($10^{4-7}$, Kennel \& Coroniti \cite{kennel84}), at the termination shock the bulk motion energy is converted into disordered (thermal) velocities, thus inside PWNe we have a relativistically hot plasma $p/\rho c^2 \gg 1$ (with $p$ thermal pressure and $\rho$ rest mass density). We adopt a value $p/\rho c^2 = 20$ which is high enough to neglect the inertia associated with the rest mass of the fluid (this is also a typical value in global simulations of PWNe, where the Lorentz factor in the unshocked pulsar wind is $\sim 100$). We use an ideal gas equation of state with adiabatic coefficient $\Gamma=4/3$, appropriate for a relativistically hot plasma. Different values of $B_z$ are considered, in order to model different intensities of the transverse magnetic field (in the following we will refer to this component of the magnetic field as ``toroidal'', $B_\mathrm{tor}$ in analogy with 2D global PWNe simulations). In addition to this main field, we also investigate cases with a (usually much weaker) magnetic field component aligned to the interface, $B_x$ (in the following we will refer to the component of the magnetic field in the $x-y$ plane as the ``poloidal'' field, $B_\mathrm{pol}$, again in analogy with the PWN case). Here we will measure the relative importance of the magnetic fields by giving the values of the non-dimensional toroidal and poloidal magnetization parameters 
\be
\sigma_\mathrm{tor}=\frac{B^2_z}{8\pi p},~~~~
\sigma_\mathrm{pol}=\frac{B^2_x}{8\pi p}.
\ee
Since in our configuration $B_z$ is the main component, $\sigma_\mathrm{tor}\ll 1$ means a pressure dominated plasma, $\sigma_\mathrm{tor}\gg 1$ means a magnetically dominated plasma, and the condition of equipartition is $\sigma_\mathrm{tor}= 1$, while we will always take $\sigma_\mathrm{pol}\ll 1$. 

The shear velocity profile in the initial unperturbed configuration is assumed to be:
\be
V_x=-V\zr\tanh{(y/\alpha)}
\ee
where the value of $\alpha$ has been chosen in order to avoid the growth of perturbation on the same scale of the cell size, which may occur if a sharp transition is used. In fact, in the limit of $\alpha\rightarrow \infty$, the system becomes scale free (Ferrari et al. \cite{ferrari80}), and given that the growth rate is proportional to the wave number, numerical noise can be strongly amplified. The main reason to introduce a shear layer of finite thickness is exactly to prevent the growth of such modes. At time $t=0$ we also impose a transverse monochromatic perturbation $V_y$ of the form:
\be
V_y=W\zr\sin{(k_x x)}\exp{[-(y/\beta)^2]}
\ee
where $W\zr$  ($\ll V\zr$) is the amplitude of the perturbation, $k_x$ is its wave number and $\beta$ a parameter that describes its fading along the $y$-direction.

In the $x$-direction the size of the computational domain is taken $D=2\pi/k_x$, and periodic boundary conditions are assumed, whereas in the y-direction boundaries are placed at larger distance, $\sim \pm 14 D$, and 0-th order extrapolation is assumed in order to minimize possible spurious reflections. We use a uniform resolution with 180 zones in the $x$-direction, while in the $y$-direction a stretched grid is employed with a uniform resolution of 360 zones in the range $[-D, D]$. In all our simulations we have chosen $\alpha=D/100$, $\beta=D/10$, and $W\zr=0.01V\zr$. The value of $\alpha$ has been chosen to be the smallest possible in order not to affect the growth rate. In all simulations we have considered monochromatic perturbations with the longest possible wavelength $\lambda=D$, for numerical convenience. From now on we normalize all lengths by this quantity, so that for example the $x$ domain is $[-1,1]$ and $k_x x = 2\pi x$. The chosen time unit is the light travel time $D/c$, so that all velocities will be expressed in units of $c$.

\section{Flow dynamics}

We will discuss here the dynamical evolution of the fluid structure and the magnetic field in our simulations. Given that our simulations are basically the relativistic extension of those of Malagoli et al. (\cite{malagoli96}), our discussion will share many common points with that work. In this paper we mainly focus on the subsonic and marginally supersonic cases. Three different values for the shear velocity are considered: $V\zr=0.1$, typical of the flow speed in the outer edge of PWNe; $V\zr=0.25$, which corresponds to a case with Mach number $\approx 1$; $V\zr=0.4$, a typical situation encountered in the inner high velocity flow channels. Concerning the toroidal magnetic field, we consider cases with $\sigma_\mathrm{tor}=0.2$ (pressure dominated regime), $\sigma_\mathrm{tor}=1$ (equipartition), and $\sigma_\mathrm{tor}=2$ (magnetically dominated regime). Different values of the poloidal magnetic field $B_x$ are then adopted to investigate its importance in the stabilization of the interface.

\begin{figure}
\resizebox{\hsize}{!}{\includegraphics{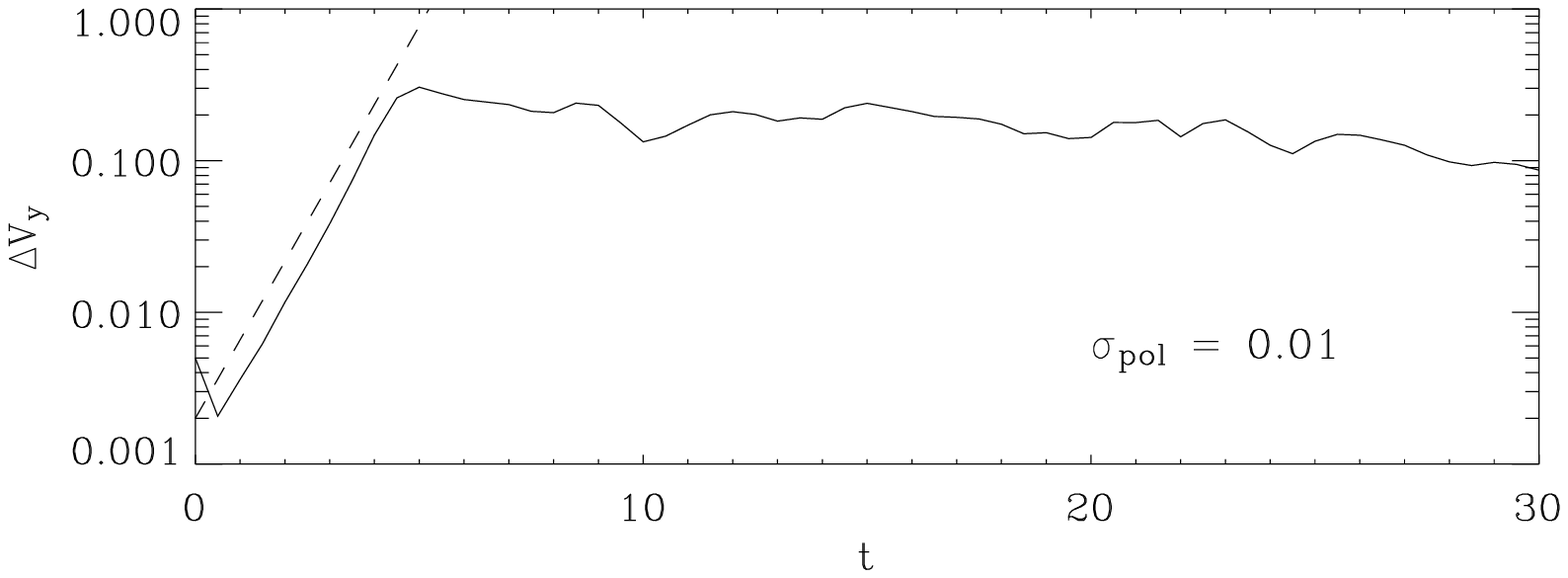}}
\resizebox{\hsize}{!}{\includegraphics{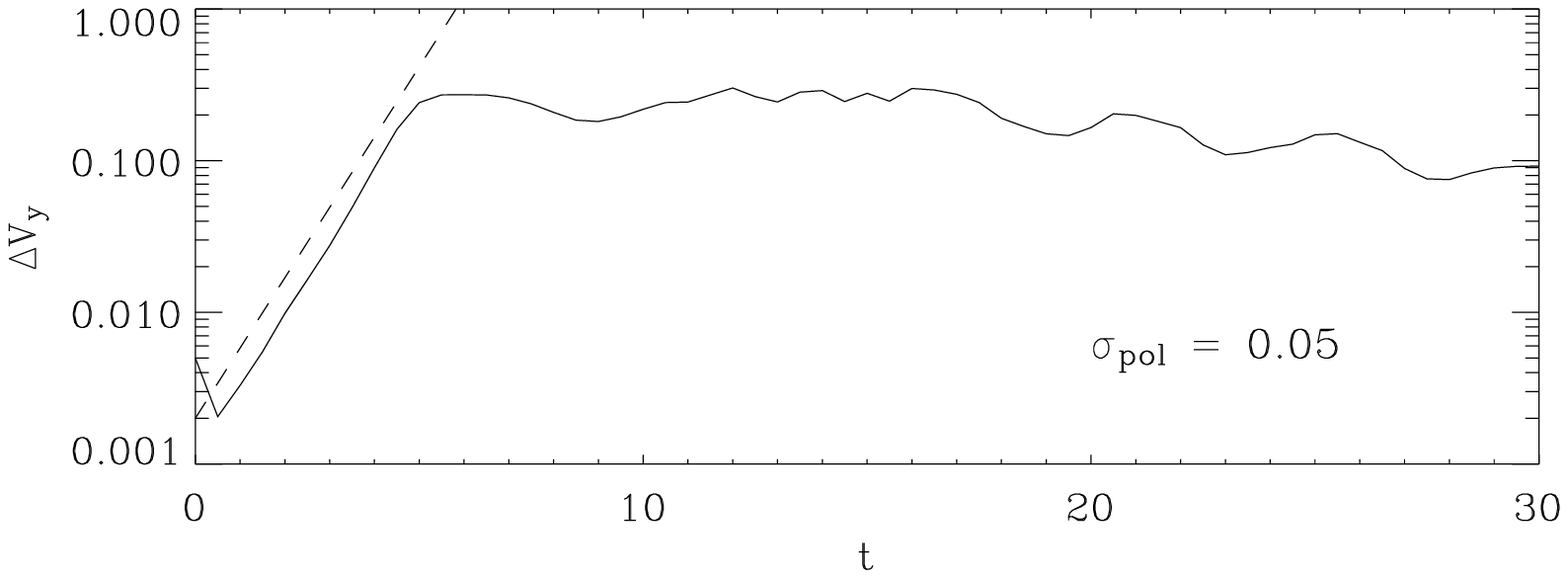}}
\resizebox{\hsize}{!}{\includegraphics{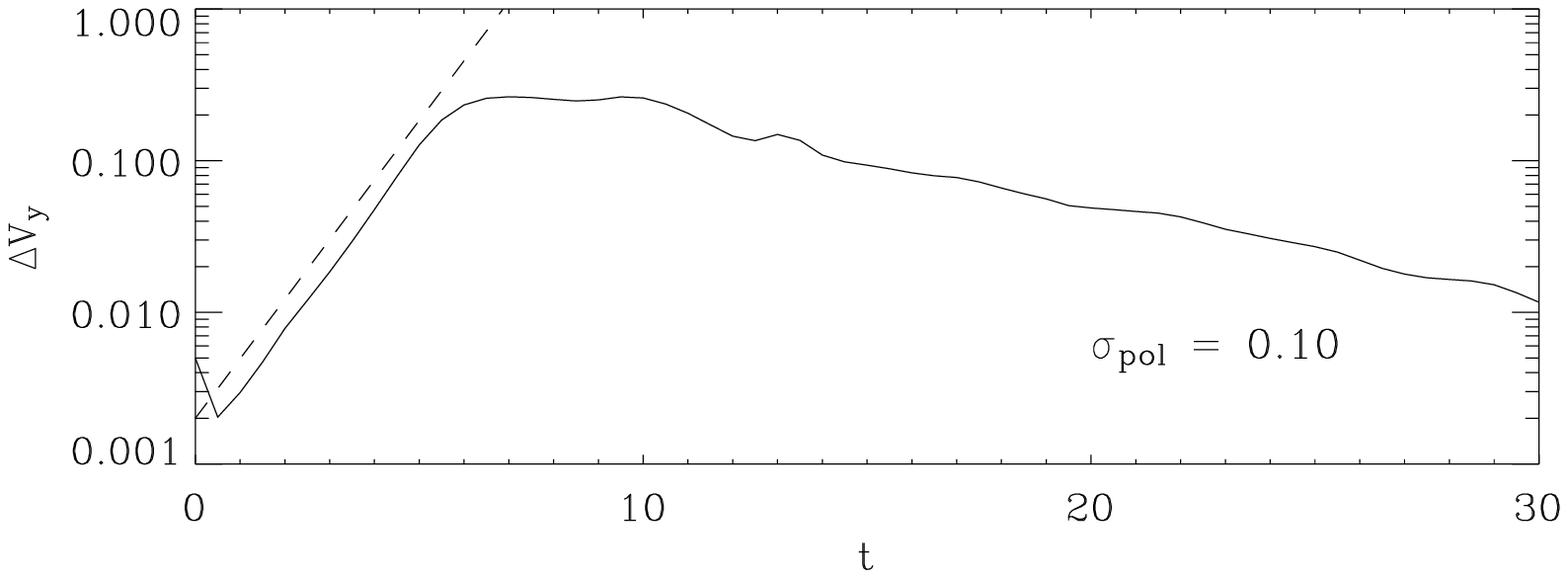}}
\resizebox{\hsize}{!}{\includegraphics{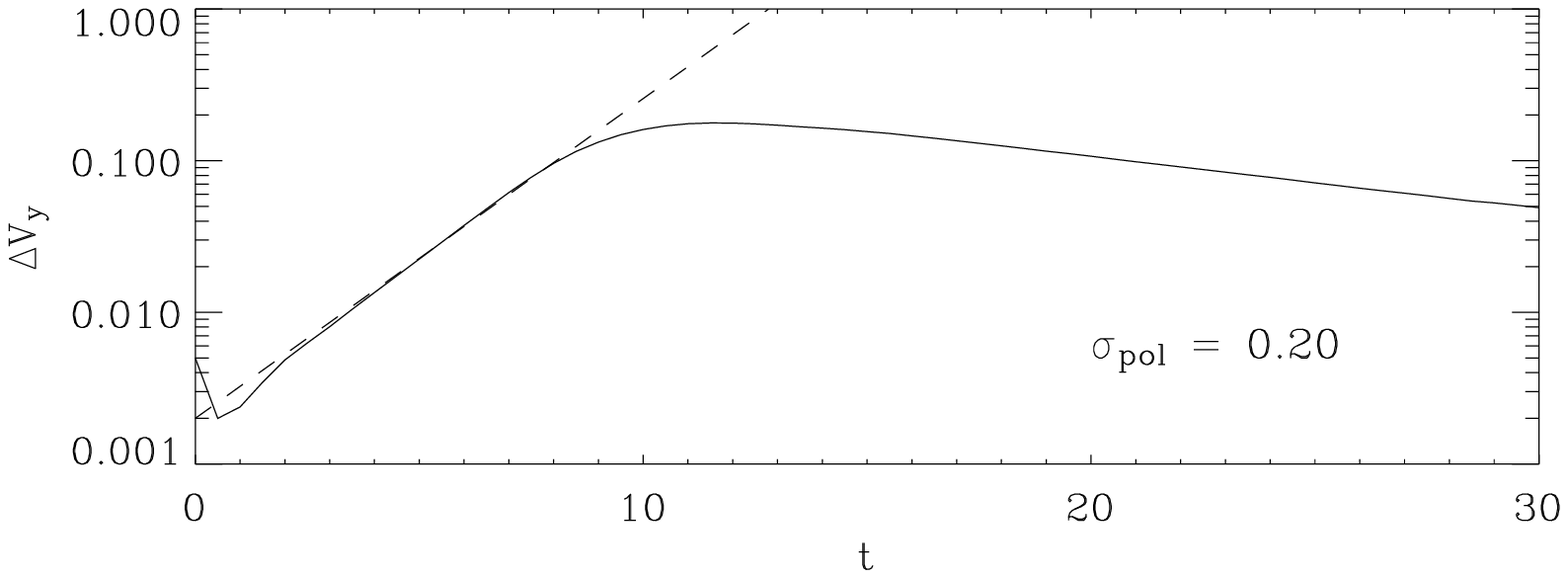}}
\caption{Evolution of the amplitude of the perturbation $\Delta V_y=0.5(V_{y,max}-V_{y,min})$ as a function of time ($V\zr=0.25$; $\sigma_\mathrm{tor}=0$). Continuous line: numerical result; dashed line: prediction of the analytic theory (Ferrari et al. \cite{ferrari80}). From top to bottom: $\sigma_\mathrm{pol}=0.01,\; 0.05,\; 0.1,\; 0.2$. The growth rates found from our numerical simulations are, respectively: 0.21, 0.11, 0.16, 0.08; to be compared with the expectations of the linear theory: 0.205, 0.108, 0.165, and 0.075.} Time is in units of $D/c$.
\label{fig:grt}
\end{figure}

During the initial phase the evolution of the perturbations follows with good accuracy the prediction of the linear theory by Ferrari et al. (\cite{ferrari80}), as shown in Fig.~(\ref{fig:grt}) where the exponential growth is well reproduced. To allow a more straightforward comparison with the analytic theory, where the growth rate is defined in a system a rest with one of the two fluids, we consider here the particular case $B_z=0$, corresponding to their Fig.~1(b) for a relativistically hot plasma\footnote{In the notation by Ferrari et al. (\cite{ferrari80}) the corresponding adiabatic factor is $\Gamma=1$, the Mach number is $M=\beta/c_s=\sqrt{3}\beta$, $\beta=2V\zr$, and the plotted growth rates must be corrected by the factors $c_s=1/\sqrt{3}$ and $\gamma_0^{-2}=1-V_0^2$}. In this simple case the value of the magnetic field that must be used in the dispersion relation does not change from the fluid to the perturbation frame, thus allowing an easier comparison. The good agreement suggests that the value we have chosen for $\alpha$ is small enough that the growth rate is not affected. In general we have verified that the growth rate tends to be smaller than the prediction of the vortex sheet approximation for larger values of $\alpha$, and that the discrepancies increase for larger shear layer. Such effect was already noticed by Perucho et al. (\cite{perucho04}) and can be easily understood in terms of stabilization by the shear layer itself. In general we find that the growth rates derived from our numerical simulations agree with the linear theory (both in the case of poloidal and toroidal magnetic field) within a few percent.

In the cases with $B_x=0$, corresponding to the assumption that inside the PWN the magnetic field is purely toroidal, it can be shown that transverse RMHD basically behaves as hydrodynamics (e.g. Romero et al. \cite{romero05}). Only pressure-like effects survive, while magnetic tension (related to the Alfv\'enic modes) is clearly absent. We find in fact that the development of the KH instability and its saturation show only a minor dependence on the magnetization of the fluid (except the change in effective sound speed and compressibility). The system is seen to evolve to form a single vortex (not shown here), rather stable, on the scale of the computational box. This behavior is typical in hydrodynamics, where vorticity evolves to the largest possible scales, although here we cannot see a proper inverse cascade since we have initialized our system right from the start with the largest possible wavelength. A pressure and density depression forms rapidly at the center of the vortex. The variation of these quantities is a function of the shear velocity: for $V\zr=0.1$, $0.25$, and $0.4$ the pressure saturates to values respectively $10$, $50$, and $75\%$ lower than in the initial unperturbed state. The transverse velocity $V_y$ grows from $W\zr$ to value $\approx\pm V\zr$, while the perturbation on the parallel velocity $V_x$ is somewhat smaller and $\approx 0.7 V\zr$. The time at which saturation is reached depends on the value of the shear velocity $V\zr$ too. In particular, the higher $V\zr$ the less it takes to saturate: for $V\zr=0.1$  the saturation time is $9$, for $V\zr=0.25$ it is $5$, and for $V\zr=0.4$ goes down to about $3-4$. On the other hand, the value of the saturation time does not seem to depend on the magnetization parameter $\sigma_\mathrm{tor}$.

\begin{figure*}
\resizebox{\hsize}{!}{\includegraphics{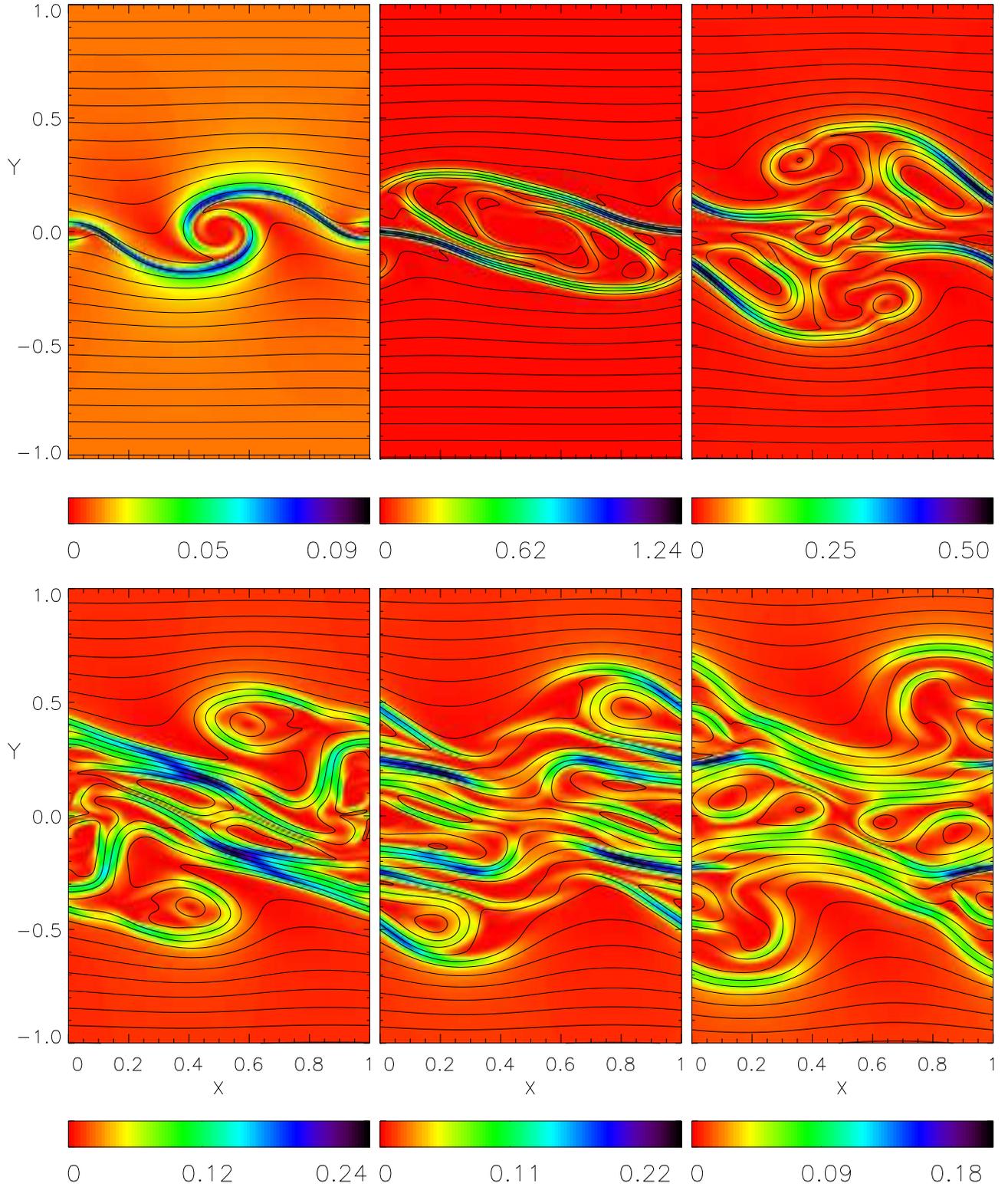}}
\caption{Evolution of the magnetic structure ($V\zr=0.25$; $\sigma_\mathrm{tor}=1$; $\sigma_\mathrm{pol}=0.01$). The contours represent poloidal magnetic field lines while the colors indicate the value of $B_\mathrm{pol}/B_\mathrm{tor}$. The wavelength of the initial perturbation is $\lambda=1$. From top to bottom and from left to right panels are uniformely spaced in time from 5 to 30. The first panel corresponds to the end of the linear phase. The second panel corresponds to the maximum amplification of the poloidal magnetic field before reconnection starts. Notice, at later times, the development of the filamentary features discussed in the text.}
\label{fig:1}
\end{figure*}

\begin{figure*}
\resizebox{\hsize}{!}{\includegraphics{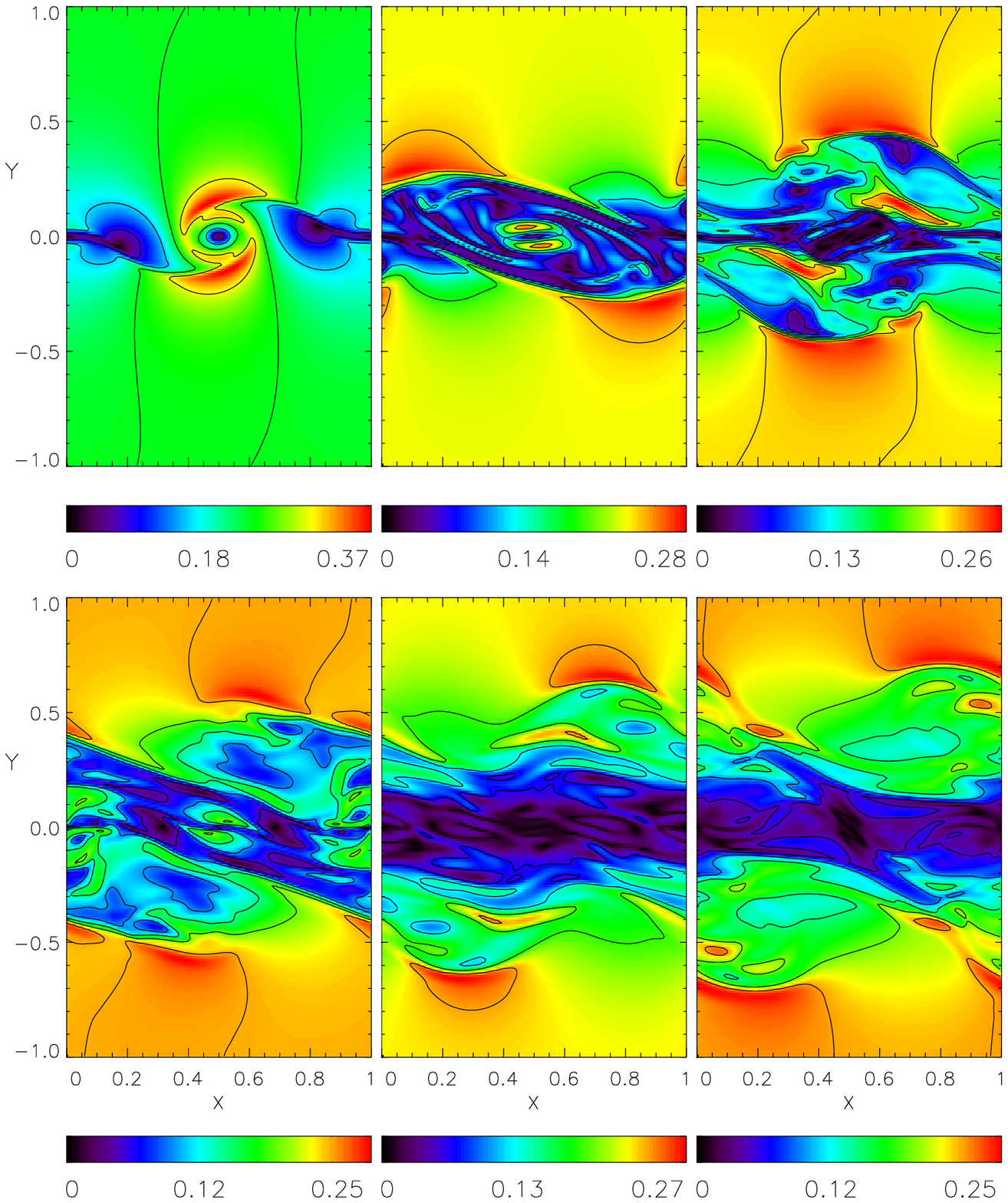}}
\caption{Evolution of the poloidal velocity for the same run ($V\zr=0.25$; $\sigma_\mathrm{tor}=1$; $\sigma_\mathrm{pol}=0.01$) and at the same times as in Fig.~\ref{fig:1}. The maximum in the value of the perturbations in the poloidal velocity is reached at the end of the linear phase, first panel. As the system evolves in time, a slow velocity central channel is formed, characterized by a filamentary wave pattern.}
\label{fig:2}
\end{figure*}

As expected, in the presence of an additional (small) poloidal magnetic field component the development of the instability changes qualitatively and there is an excitation of small scale turbulence. As a reference case we will consider here the case $V\zr=0.25$. The only major quantitative differences with the cases with higher and lower shear velocity is that the stabilization due to a poloidal magnetic field is more efficient the lower the value of $V\zr$. However, even in the case $V\zr=0.4$ a magnetic field such that $\sigma_\mathrm{pol}=0.1$ is sufficient to strongly suppress the KH growth. The linear phase of the growth in the presence of a weak poloidal magnetic field ($\sigma_\mathrm{pol}=0.01$), is essentially the same as in the hydrodynamical case (see Fig.~\ref{fig:1}). The magnetic field frozen into the fluid gets stretched and twisted as the large scale vortex develops. During the linear phase the amplitude of the perturbation in the velocity field is approximatively the same as in the hydrodynamic (or transverse magnetohydrodynamic) case. Looking at Fig.~\ref{fig:2} we observe that in the linear phase the poloidal velocity can reach values as high as 0.4 (to be compared with $V\zr=0.25$). However, the decay of the MHD instability strongly reduces the perturbation and at the center of the shear layer the poloidal velocity drops to value close to 0. On the contrary the density and pressure perturbations are strongly reduced with respect to the hydrodynamical case, to values not exceeding 25\%.

During the linear phase the poloidal magnetic field is amplified to values of order $\sqrt{4pV\zr^2}$ (equipartition with the kinetic shear energy). The amplification does not seem to depend on the value of the toroidal magnetic field, probably because in our cases the enthalpy is pressure dominated. The other interesting property is that the poloidal magnetic field perturbation reaches a maximum after the time at which the velocity perturbation peaks, with time scales longer of a factor 2, approximately. This is probably related to numerical resistivity effects. It is know in fact that the higher the resistivity the lower is the amplification of the poloidal field. We want also to point out that in our 2.5D approximation such magnetic amplification is only a transient effect, because in our geometry real dynamo processes are not possible. As we can seen from Fig.~\ref{fig:1} the magnetic field is stretched by vortical flows, the gradient scales get smaller, until the resistive scale is reached. This is the moment when reconnection of the poloidal magnetic field starts and represents the end of the linear phase. This reconnection process takes place in an intermittent manner. After the first reconnection, the magnetic field is again twisted and wound up until another reconnection event occurs. Such behavior is also reflected in the velocity perturbations. After the first reconnection, the velocity perturbation is allowed to grow again. This phase is also characterized by an increase in the extent of the mixing layer. 

\begin{figure*}
\resizebox{\hsize}{!}{\includegraphics{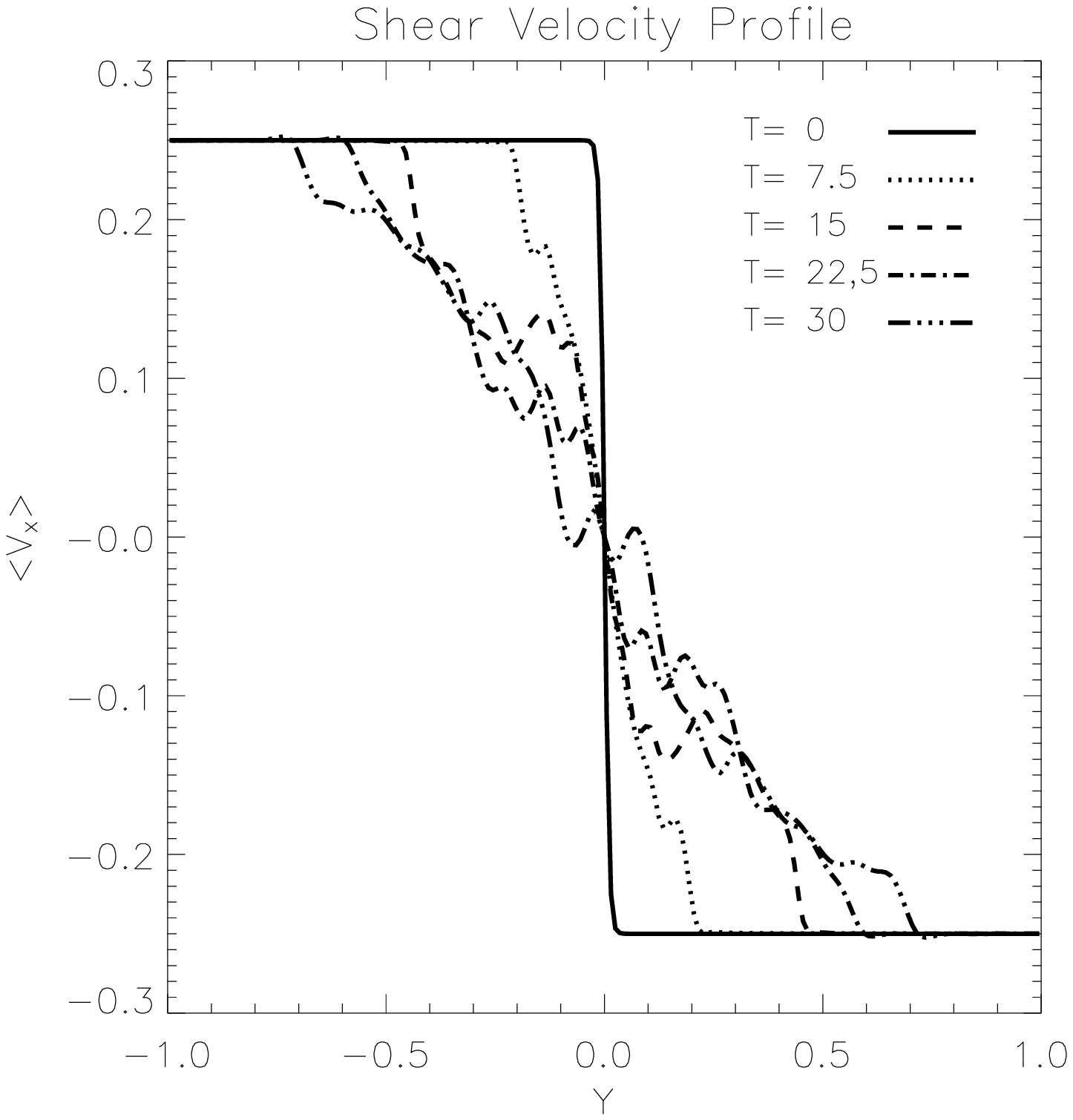}\includegraphics{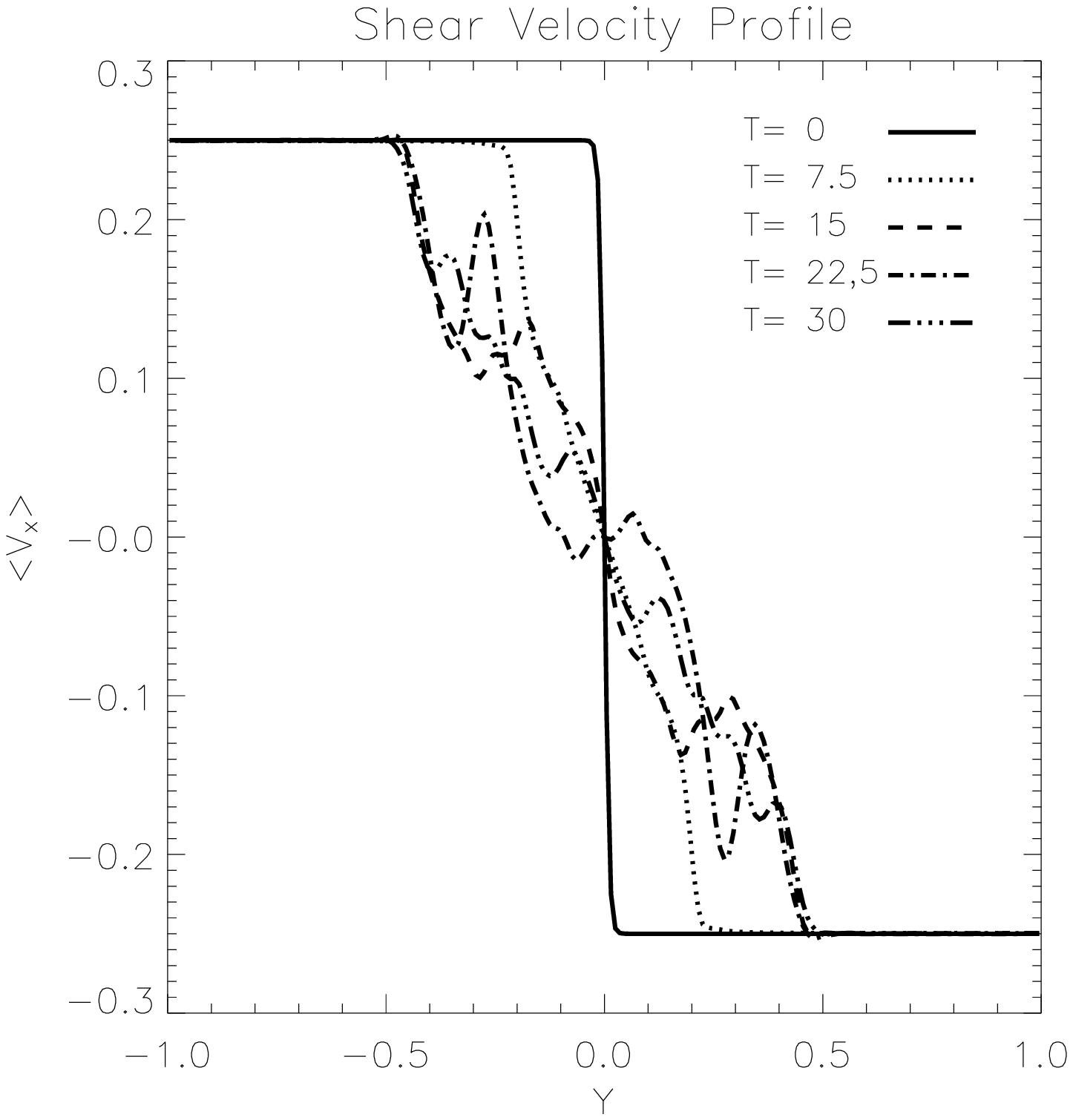}
\includegraphics{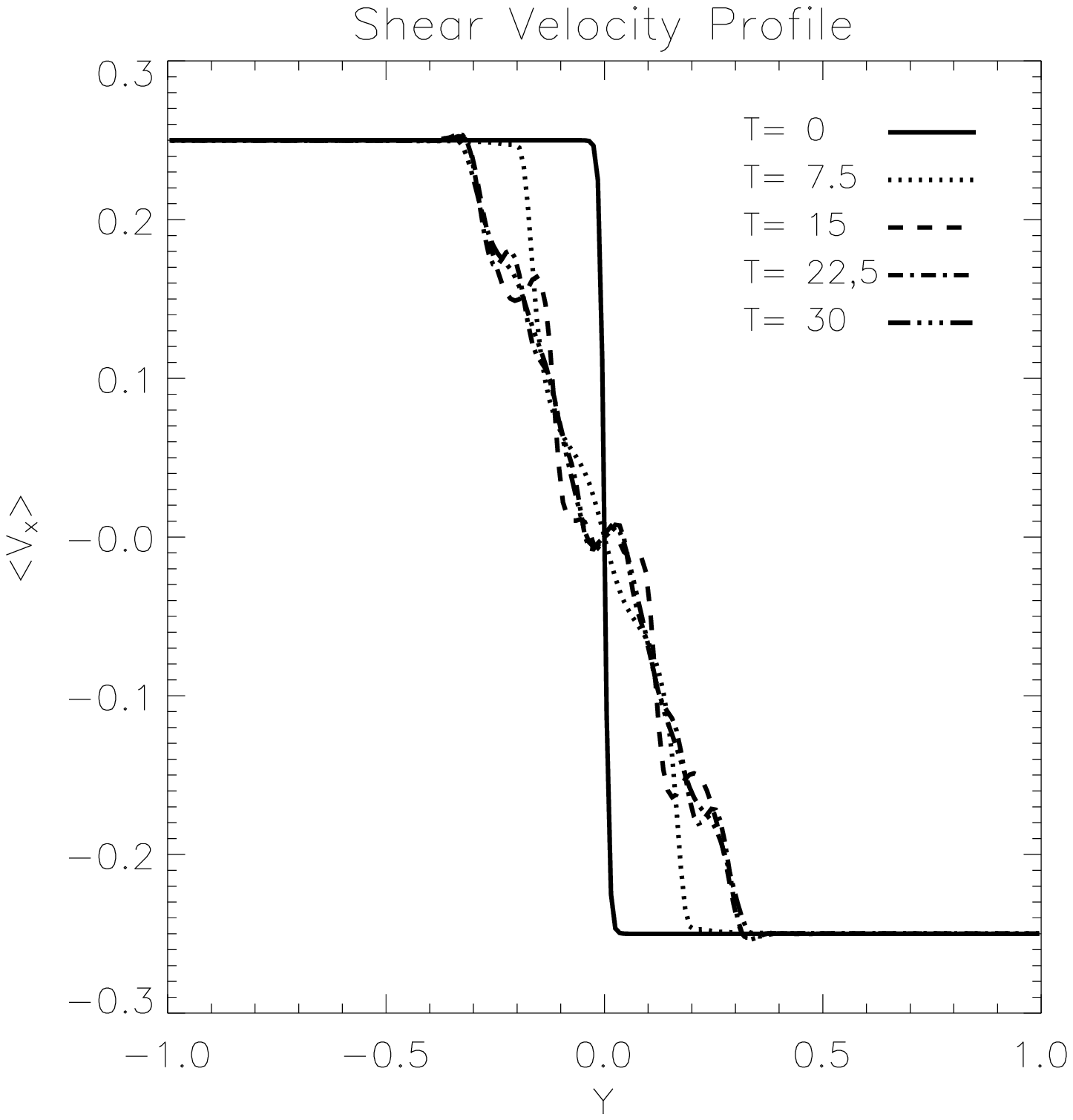}}
\caption{Evolution of the integrated velocity profile across the shear layer ($V\zr=0.25$, $\sigma_\mathrm{tor}=1$). From left to right: $\sigma_\mathrm{pol}=0.01$, $\sigma_\mathrm{pol}=0.05$, and $\sigma_\mathrm{pol}=0.1$. Note that for $t>15$ the profile is relaxed to an almost steady structure, in all cases. For stronger poloidal fields the velocity profile evolves to narrower and smoother transitions, typical of the stabilizing effect of the magnetic field parallel to the shear layer.}
\label{fig:3}
\end{figure*}

As the magnetic turbulence decays in this direct cascade to smaller scales a new statistically steady flow is reached. A large mixing layer is formed, whose  size depends on the value of the shear velocity and is of order of the size of the original perturbation. The mixing layer is characterized by filamentary structures, which are known features of decaying MHD turbulence, approximatively aligned with the direction of the unperturbed interface. In comparison with the corresponding hydrodynamical case the extent of the shearing layer is about twice as bigger. Looking at Fig.~\ref{fig:1} we notice that the poloidal magnetic field is strongly amplified in those filamentary structures (which, as pointed by Malagoli et al. \cite{malagoli96}, turn out to be slow magnetosonic waves) to values that, at the end of the linear phase, can exceed the amount of toroidal magnetic field. However, in the equipartition case $\sigma_\mathrm{tor}=1$, the poloidal field compressed in the filaments tends to saturate, to values from 10\% of the toroidal field for $V\zr=0.1$, up to 40\% for $V\zr=0.4$ (apparently with a linear scaling). In the case of stronger field, the system tends to evolve to a stable configuration where the shear layer is characterized by a series of filamentary structures that are slightly inclined with respect to direction of the initial interface. 

\begin{figure}
\includegraphics[width=7cm]{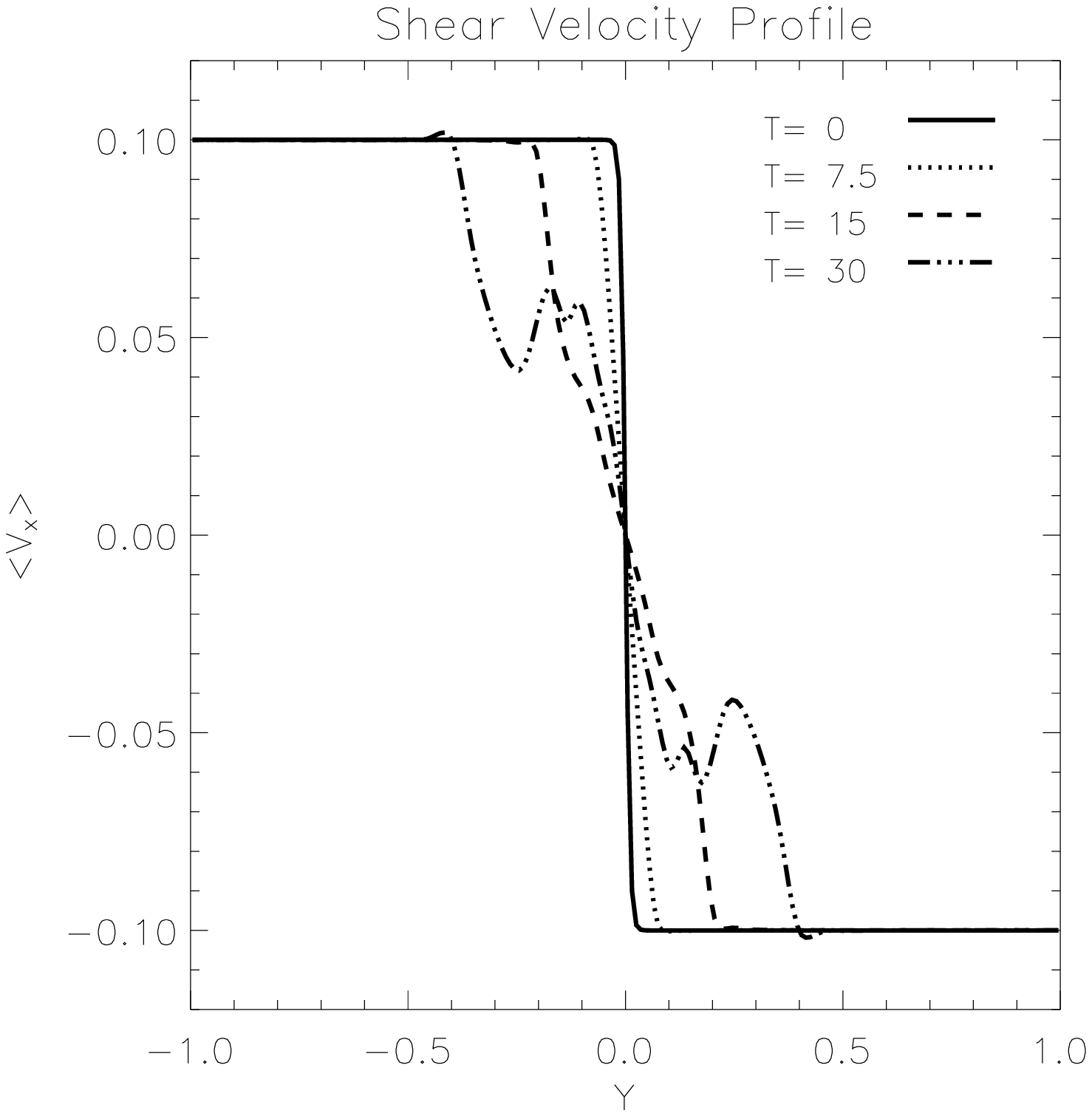}
\includegraphics[width=7cm]{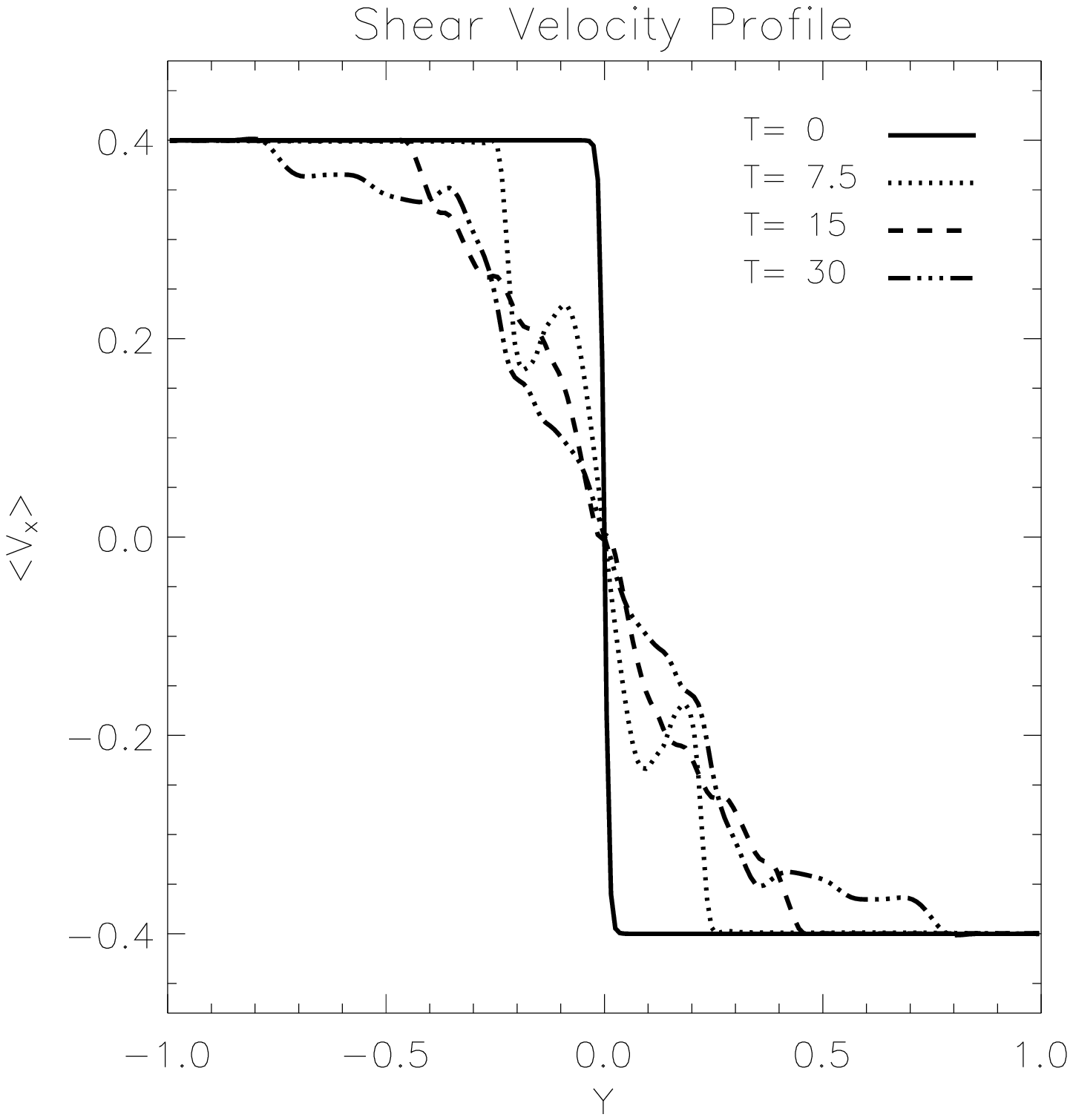}
\caption{Evolution of the integrated velocity profile across the shear layer ($\sigma_\mathrm{tor}=1$, $\sigma_\mathrm{pol}=0.01$). Top panel $V\zr=0.1$, bottom panel $V\zr=0.4$. Notice that for $V\zr=0.1$ the extent of the shear layer is smaller than in the case $V\zr=0.25$ (Fig.~\ref{fig:3}), while no major differences are shown in the $V\zr=0.4$ case.}
\label{fig:4}
\end{figure}

If we now follow the evolution of the shear layer width, we notice that it increases during the linear and subsequent reconnection phase, and then saturates at later times. The degree of mixing inside the layer is complete in the case of a weak poloidal field, although there is no appreciable mass loss from the layer. This implies that if KH instability is at work in PWNe, the amplified small scale magnetic fields tend to be confined in the vicinity of the shear layers. Moreover the extension of the mixing layer (evaluated by using a passively advected scalar) turns out to be smaller (about one half) than the size of the shear layer as defined by the $V_x$ profile, averaged in the $x$ direction (Fig.~\ref{fig:3}-\ref{fig:4}). With a stronger poloidal field, the growth of the shear layer is not accompanied by any mixing. Apparently the KH instability is able to amplify the poloidal magnetic field on timescales of order of the saturation time. Thus, even if, in the simplified geometry of out simulations, permanent dynamo processes are prevented, the decay of the amplified magnetic field appears to be very long. 

The amplification of the poloidal magnetic field at shear layers may be very important on the observational side as well. We recall that, during the linear phase, the field components increased by the instability tend to reach equipartition with the flow itself, thus for stronger velocity shears a higher poloidal field is expected. As pointed out in the introduction, a purely toroidal field seems unable to explain emission features like the jet. However, material flowing into the jet comes from high velocity channels, diverted toward the axis by hoop stresses in a sort of back-flow. In the case of a striped wind, that is when we have a region with vanishing toroidal field around the equator, the shear of such channels is against an equatorial (unmagnetized) flow, which is not inhibited as in the case of full magnetization and that can be actually quite strong (Del Zanna et al. \cite{ldz04}). Therefore, the resulting shear layer may contribute greatly to the amplification of a small scale poloidal magnetic field, which can then in part be advected with the fluid toward the axis and thus be responsible for the observed polar emission.

\section{Synchrotron emission}

It is evident from the above discussion on the dynamics arising from the development of KH instability, that the resulting vortical motions and the amplification of poloidal magnetic field lead to substantial deviations from the laminar unperturbed flow. It is natural to expect such modifications to be reflected in the emission properties and to provide a possible explanation for the observed time variability. Following Bucciantini et al. (\cite{bucciantini05a}), to which the reader is referred for a derivation of the equations, the local synchrotron emissivity $s(\nu)$ at a frequency $\nu$ can be approximated by:
\be
s(\nu)\propto p D^{2+\xi}(B'_\perp)^{1+\xi}\nu^{-\xi},
\label{eq:snu}
\ee   
where $p$ is the thermal pressure, $2\xi +1$ is the power law index of the particle distribution (in the following we have adopted the value $\xi=0.6$), $B'_\perp$ is the magnetic field component perpendicular to the line of sight, as measured in the fluid frame, and $D=[\gamma(1-\vec{\beta}\cdot\vec{n})]^{-1}$. Here $\vec{\beta}$ is the flow velocity and $\vec{n}$ is the versor pointing toward the observer. Given the magnetic field structure one can also compute the polarization fraction and the polarization angle. The value of the Stokes parameter $U$ and $V$ is derived following Lyutikov et al. (\cite{maxim}) (see also Bucciantini et al. \cite{bucciantini05b}, Blandford \& Konigl \cite{blandford79}). To correctly evaluate the emission properties it is fundamental that the flow structure is computed using relativistic fluid dynamics, and the Doppler term $D$ may become one of the most important ingredients.

Looking at Eq.~\ref{eq:snu} it is evident that there are two main source for modulation of the synchrotron emission. The first is the compression or rarefaction connected with the development of vortexes, which reflects itself in the enhancement or decrement of the $p$ and $B'_\perp$ terms. In the cases with a purely toroidal magnetic field the emissivity in the center of the largest scale vortex can be strongly suppressed with respect to the surrounding laminar flow. In the case $V\zr=0.4$, for example, the central emissivity is about 10 times lower than in the unperturbed regions. However, in the presence of a poloidal field, the vorticity decays to smaller scales (the large scale eddy that forms during the linear phase is disrupted, and the resulting turbulent shear layer contains many eddies with sizes smaller thn the original perturbation) and such compressible modulation is strongly reduced. The second source of modulation is due to the boosting term $D$, which amplifies the emission from a turbulent fluid at points where the local velocity is toward the observer. Even in the presence of a weak poloidal field, the transverse velocity $V_y$ can reach values comparable with $V\zr$. The boosting term can account for fluctuations in the emissivity of about a factor 10 for $V\zr=0.4$ and a factor 4 for $V\zr=0.25$. Finally, as we can see from Fig.~\ref{fig:1}, the poloidal magnetic field can be amplified to a non negligible fraction of the toroidal field. A poloidal field on small scales, while still contributing to the emissivity, might lead to a depolarization of the observed radiation (if the field is purely toroidal the polarization is by definition at its theoretical maximum).  

\begin{figure*}
\resizebox{\hsize}{!}{\includegraphics{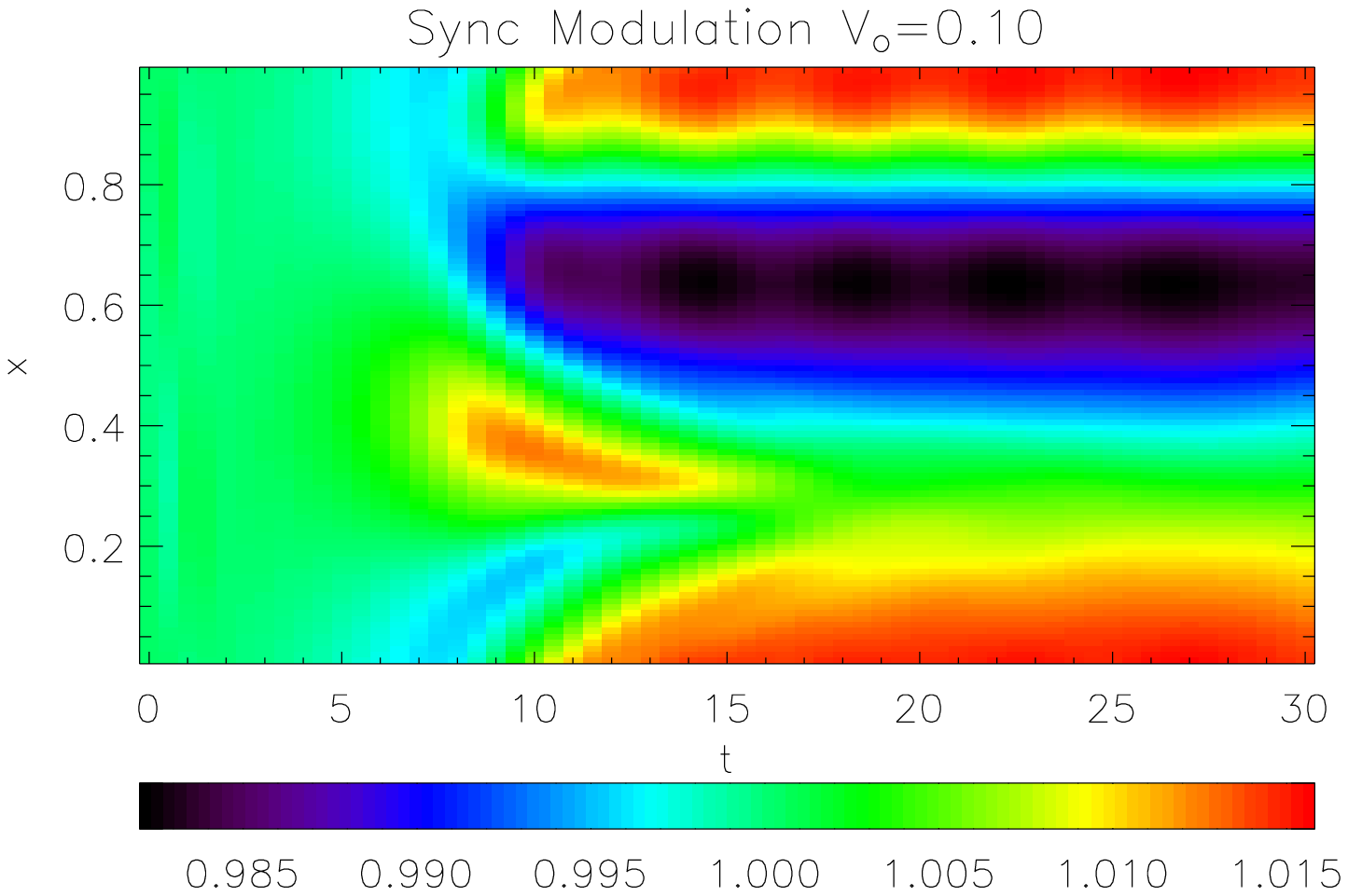}\includegraphics{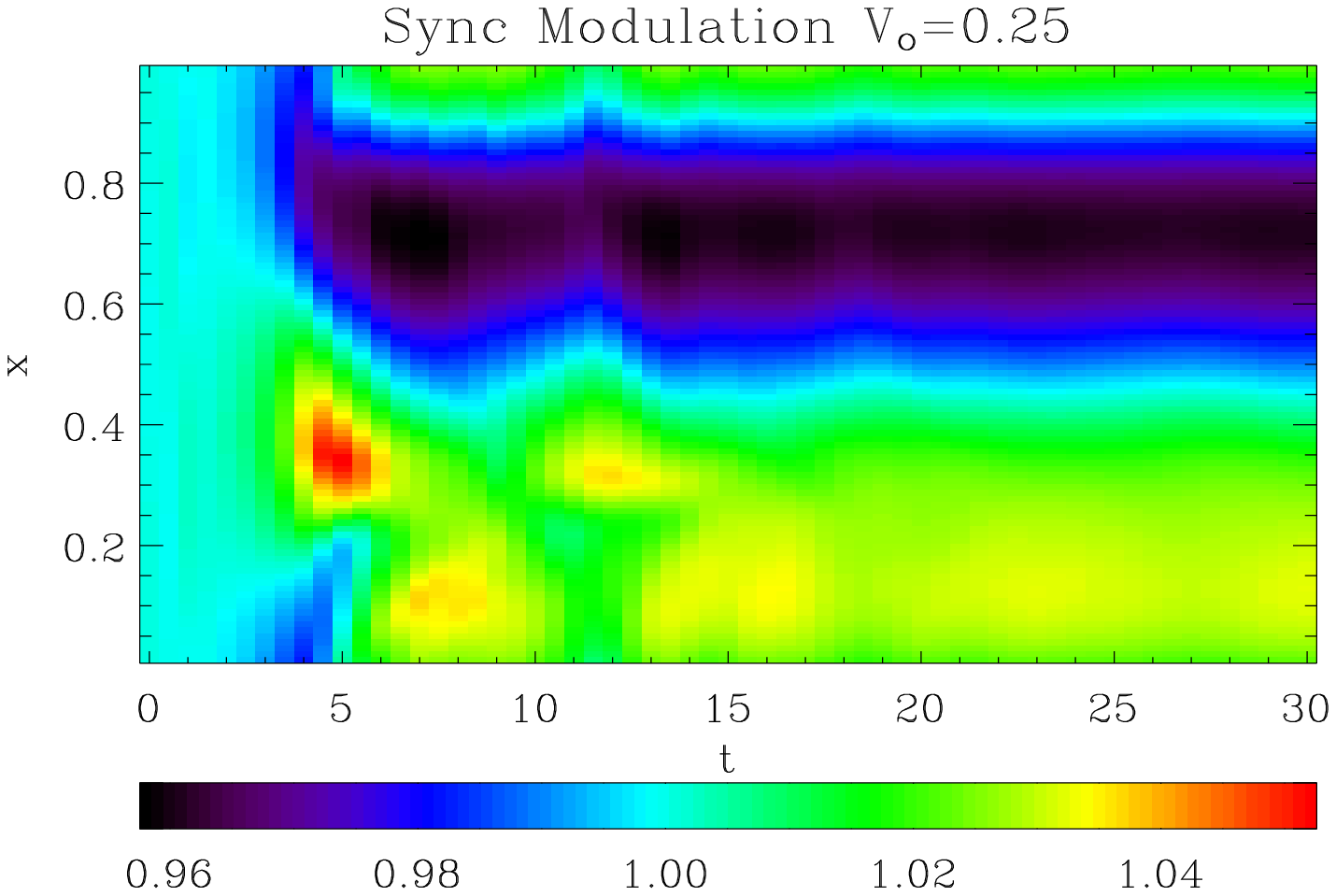}\includegraphics{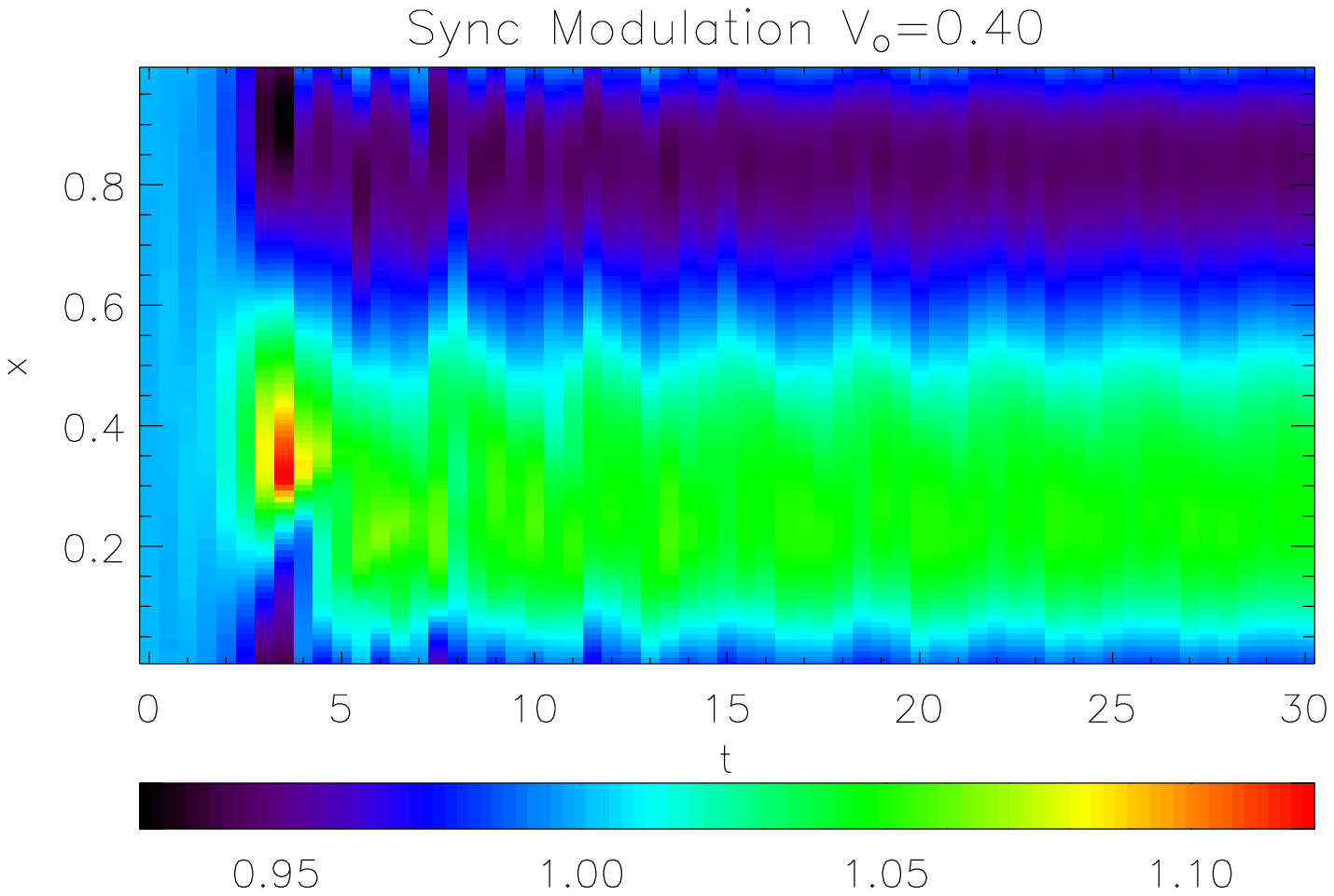}}
\caption{Time evolution of the synchrotron emission, normalized to the mean value, for an inclination angle of $30^\circ$. A purely toroidal magnetic field at equipartition $\sigma_\mathrm{tor}= 1$ is assumed in the unperturbed state. From left to right: $V\zr=0.1$,  $V\zr=0.25$ and  $V\zr=0.4$. }
\label{fig:5}
\end{figure*}

However, in spite of the local fluctuations in the emissivity, associated with the growth of the KH instability, which could in principle account for large modulation in the synchrotron emission, it is not obvious how, once the contribution from different fluid elements is integrated along the line of sight, the resulting modulation will actually look like. Moreover, in the case of the Crab Nebula, the inclination of the equatorial plane (which should be parallel to the shear interface of the flow channels) and the observer is about $30^\circ$ (Weisskopf et al \cite{weiss00}). Once the contribution of the various fluid elements along the line of sight is integrated, the combined effect might lead to a suppression of the synchrotron modulation with respect to what could be inferred by just considering local conditions. To compute the total emission, the integration along the line of sight was performed on a layer extended in the range $-1<y<1$. In general, larger integration layers yield smaller modulations. Such size was chosen to guarantee that in all cases the shear layer was contained within the integration region. For example, by considering just the contribution of the compressible term in Eq.~\ref{eq:snu}, in the case of a purely toroidal field we find that, for $V\zr=0.4$, the fluctuations in the integrated emission are about 50\% for an inclination angle of $90^\circ$, but they drop to less than 5\% for an angle of $30^\circ$. 

Let us first consider the case of purely toroidal magnetic field. Looking at Fig.~\ref{fig:5} we observe that the maximum in the synchrotron modulation coincides with the end of the linear phase ($t=5$). Notice also that at this time the synchrotron perturbation has a wavelength which is about one half that of the original perturbation. This is due to the fact that the compressible and boosting terms do not sum up in phase. The amplitude of the modulation in the synchrotron emission is however extremely small: it increases for higher $V\zr$ but in all our cases is below 10\%. At later times, in the saturation regime, the synchrotron modulation is almost completely due to the compressible terms. Once a stable vortex is established, the effects of boosting from different fluid elements tend to compensate each other. We verify that such a behavior does not depend on the assumed inclination angle. These results also show that the timescale for the onset of synchrotron modulation are of order of the saturation time.  

\begin{figure}
\resizebox{\hsize}{!}{\includegraphics{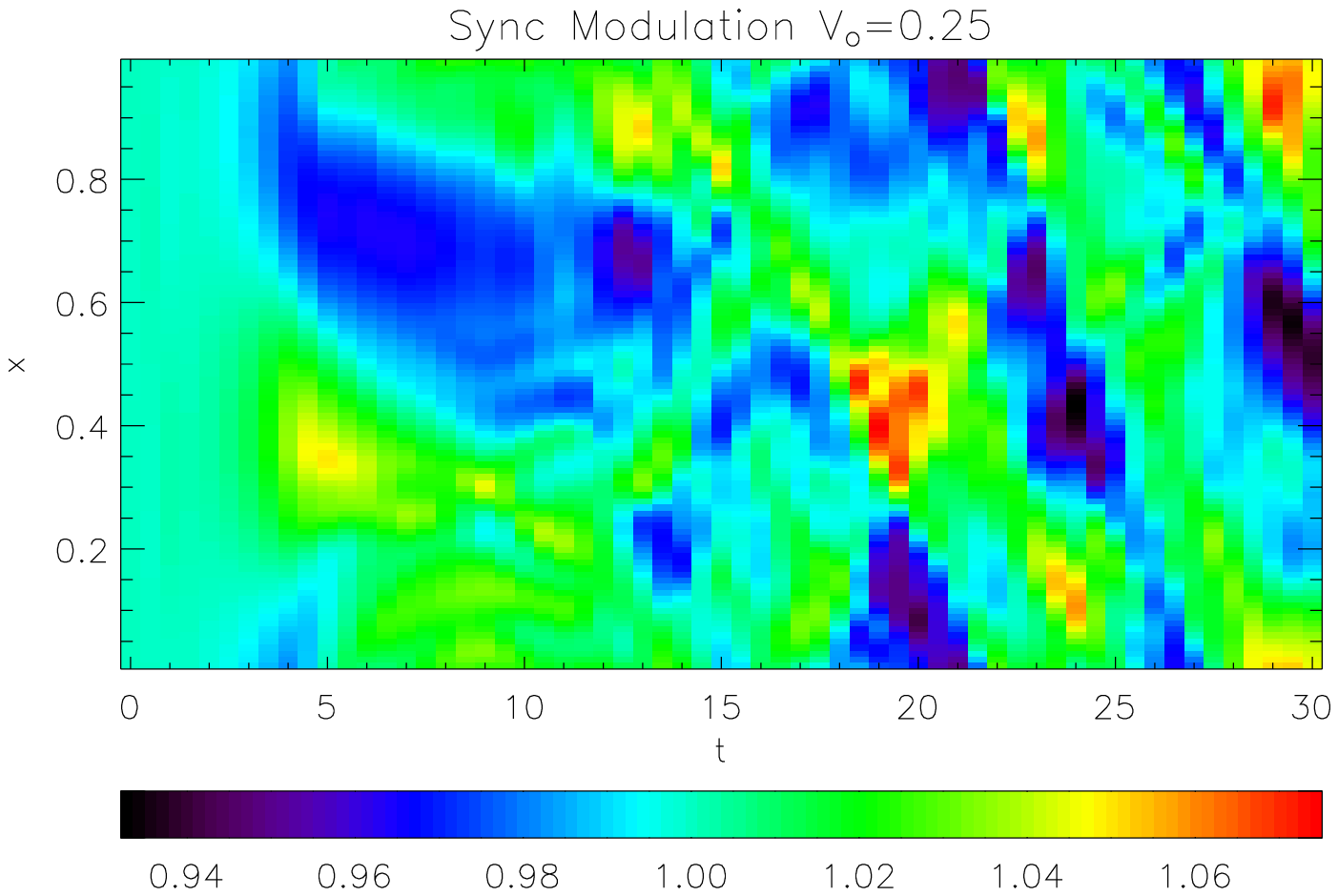}}
\resizebox{\hsize}{!}{\includegraphics{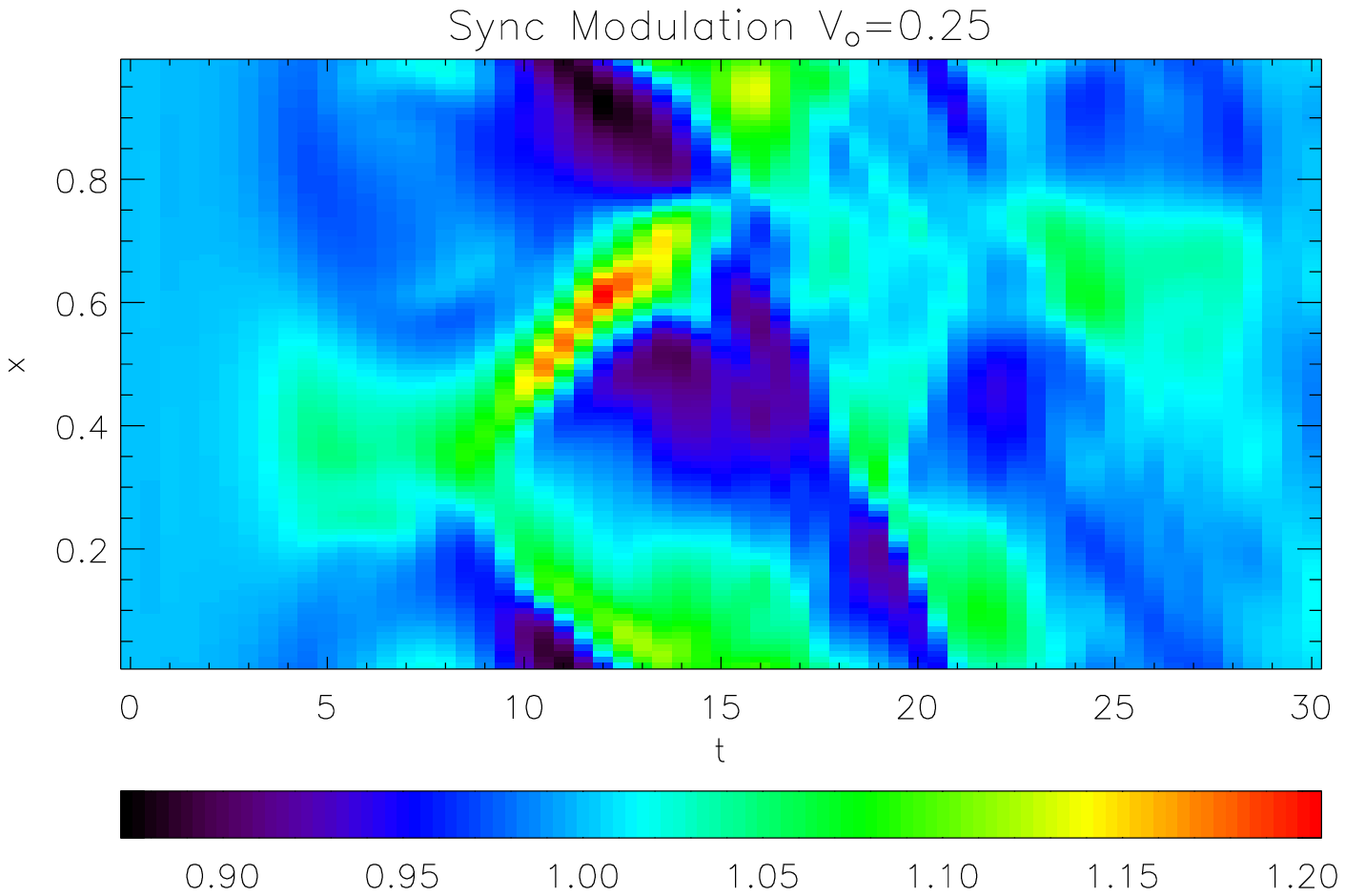}}
\caption{Time evolution of the synchrotron emission, normalized to the mean value, for an inclination angle of $30^\circ$, a shear velocity $V\zr=0.25$, a toroidal magnetic field at equipartition $\sigma_\mathrm{tor}= 1$. Upper panel: poloidal field $\sigma_\mathrm{pol}=0.01$. Lower panel: $\sigma_\mathrm{pol}=0.05$.}
\label{fig:6}
\end{figure}

If a poloidal magnetic field is present the decay of the turbulence to small scales qualitatively changes the emission signature. Until the end of the linear phase and before reconnection takes place the modulation in the synchrotron emission is the same as in the cases with purely toroidal magnetic field. After reconnection, the decay to small scales leads to a more turbulent and time dependent behavior. Looking at Fig.~\ref{fig:6} we see that the amplitude of the modulation is larger, even if still less than 10-15\%. There is also an intrinsic time variability on timescale of a few units ($D/c$). Modulations do not appear stationary in the original perturbation frame: looking at emission peaks, they appears to move with a speed $\sim 0.1c$, smaller than the shear velocity $V\zr$. This is an important difference. In the purely toroidal case (the modulation is stationary in the perturbation frame) the motion of the observed features can in principle be related to the underlying flow structure, while if a poloidal field is present such correlation is not possible. For higher values of the poloidal magnetic field the amplitude of the modulation at the end of the liner phase is higher but, as expected, the stabilization of the layer leads to a suppression at later times. In general, the modulation increases in the linear phase and is later dumped. Contrary to the purely toroidal case, where the modulation in the emission was mainly due to the compressible term, now pressure fluctuations are much smaller and the modulation in the synchrotron maps are mainly due to the boosting term. We also computed the polarization properties of the radiation, to verify if the amplified poloidal field could possibly lead to any depolarization. Despite the high value reached by $B_\mathrm{pol}/B_\mathrm{tor}$, no depolarization was observed. The reason can be understood by looking at Fig.~\ref{fig:1}: the poloidal field is amplified to high values only in thin filamentary structures. The contribution of these regions to the integrated emission is negligible, and the resulting polarization is the same as in the unperturbed state.

Let now apply the above results to the case of the Crab Nebula. Recent observations have shown intrinsic variability of the wisps region and X-ray images have shown the presence of time dependent filamentary structure in the main torus (the wavelength of the filaments is about 0.2 Ly) (Hester et al. \cite{hester02}). Recent papers published on the subject were more focused on the time variability of the wisps region, and no quantitative information about the observed synchrotron modulation is available, either for the wisps or for the torus. It is however evident from the maps that wisps are characterized by large amplitude fluctuations ($\sim 100\% $ of the wisps intensity) while the filamentary structure in the main torus is much fainter. If one considers the amplitude of the modulation, it is evident from our results that fluctuations lager than 10-20\% can be hardly due to KH instability. We thus conclude that KH is unlikely to be at the origin of the variability in the wisps, while it is still possible that it could account for the filamentary features observed in the main torus. 

Moreover, by looking at the time scale for the growth of the modulation, we find that this is too long to explain the wisps region variability (which in recent PWN models turns out to be at the termination shock). Perturbations having wavelengths of order of the main torus filaments are expected to grow with timescales of about 1-2 yr, in the linear phase. Assuming a flow velocity of order $c/3$ in the equatorial plane (derived from Doppler boosting of the torus), perturbations will have moved to a distance of about 3-6 ly, which is in rough accord with the location of the main torus (this also implies that the amplitude of the original perturbation cannot be too small). Actual data do not allow to understand if the observed time variability of modulation can be ascribed to a poloidal field, however the intrinsic time variability in our simulations is of order of years, much longer than the observed week-month timescales. It is more likely that the time variability to be due to the propagation of the perturbation, instead of its decay to smaller scales. 

The last point concerns the polarization. Optical data (Schmidt et al. \cite{sch79}) show that polarization in the wisps and main torus is about 50-70\% of the theorical maximum. Such depolarization could actually be due just to foreground and background emission. However, if we assume depolarization to be intrinsic, our result then shows that small scale magnetic fields must be injected at the shock, because amplification of poloidal field by the KH instability does not produce any appreciable depolarization. We conclude that KH instability is unlikely to be the cause of the wisps variability (it is more likely due to global modulation in the PWN flow structure, Komissarov \& Lyubarsky \cite{komissarov04}, Bogovalov et al. \cite{bogovalov05}), while it remains a good candidate for the origin of the time dependent filamentary structure in  the main torus.  

\section{Conclusions}
In this paper we have extended previous results about KH instability in the presence of a component of the magnetic field parallel to the interface (Miura \cite{miura84}, Malagoli et al. \cite{malagoli96}), to a relativistic regime typical of the conditions inside PWNe. For the first time, based on our numerical models we have derived the effect of KH instability on the emitted synchrotron radiation and compared the results with the observed time variable features in the wisps region and in the main torus of the Crab Nebula.

Regarding the evolution of the instability, we recover a similar behavior as in the non relativistic case. For a purely toroidal field the vorticity decays to larger scales (the system is seen to evolve to form a single vortex rater stabel on the scale of the computational box), transverse velocity increases to a value close to the shear velocity and pressure and magnetic field drop substantially in the vortex. The saturation time is longer for smaller shear velocity, and in the case of our initial perturbation is of order of a few $D/c$. If a small poloidal field is also present, the  evolution of the instability changes. The linear phase evolves as in the case of a purely toroidal magnetic field, until the twisted poloidal component experiences reconnection events, which take place in an intermittent manner. MHD turbulence develops and decays to the dissipative small scales set by viscosity and magnetic resistivity (in our case the numeric viscosity and resistivity of the code), these are the same scales on which reconnection takes place. Fluctuations in the pressure become negligible at this stage, while those in the transverse velocity are still high. The poloidal magnetic field is compressed in filamentary features up to values about 10-40\% of the equipartition field. There is also an increase in the extent of the turbulent shear layer.

As we have shown, fluctuations in the local emissivity due to KH instability can be very large, however once the emission is integrated along the line of sight the effects of different fluid patches tend to compensate, and only minor fluctuations remain. This stresses the importance of correctly computing the contribution from different fluid elements for the integrated emission, when comparing models to observations. Even if local variation can explain the observed features, it is not obvious how this will affect the total observed emission. We find, in our settings, that in general fluctuations do not exceed 10-20\%. In the purely toroidal case these are mainly due to compression effects on the thermal and magnetic pressure terms, while, in the presence of an initial poloidal field, modulations seem to be more related to the boosting term, which may turn on due to the turbulent velocity field produced after the saturation phase. The time scale for the growth of the synchrotron modulation is of the same order of the time scale for the instability to saturate. In the turbulent phase the time scale is still of order of few $D/c$. By comparing our results to observations of the Crab Nebula we concluded that local KH instability is unlikely to be at the origin of the wisps variability, while it remains a possible explanation for the time dependent filamentary structure seen in the main torus.

\begin{acknowledgements}
This research was supported in part by the national Science Foundation under Grant No.PHY99-07949. N.~B. was supported in part by NASA grant TM4-5000X to the University of California, Berkeley, and a David and Lucile Packward Fellowship to Eliot Quataert. 
\end{acknowledgements}


\end{document}